\documentclass[12pt]{article}
\usepackage{graphicx}
\usepackage{amssymb}
\usepackage{epstopdf}
\begin{document}

\newcommand{\overbar}[1]{{\bar{#1}}}
\newcommand{\imag}{i}
\newcommand{\Bra}[1]{{\langle{#1}|}}
\newcommand{\Ket}[1]{{|{#1}\rangle}}


\newcommand{\bb}{\begin{equation}}
\newcommand{\ee}{\end{equation}}
\newcommand{\bbb}{\begin{eqnarray}}
\newcommand{\eee}{\end{eqnarray}}
\newcommand{\diag}{\mbox{diag }}
\newcommand{\Str}{\mbox{STr }}
\newcommand{\Tr}{\mbox{Tr }}
\newcommand{\Det}{\mbox{Det }}
\newcommand{\C}[2]{{\lk [{#1},{#2}\re ]}}
\newcommand{\AC}[2]{{\lk \{{#1},{#2}\re \}}}
\newcommand{\kk}{\hspace{.5em}}
\newcommand{\vc}[1]{\mbox{$\vec{{\bf #1}}$}}
\newcommand{\mc}[1]{\mathcal{#1}}
\newcommand{\del}{\partial}
\newcommand{\lk}{\left}
\newcommand{\ave}[1]{\mbox{$\langle{#1}\rangle$}}
\newcommand{\re}{\right}
\newcommand{\pd}[1]{\frac{\del}{\del #1}}
\newcommand{\pdd}[2]{\frac{\del^2}{\del #1 \del #2}}
\newcommand{\Dd}[1]{\frac{d}{d #1}}
\newcommand{\sech}{\mbox{sech}}
\newcommand{\pref}[1]{(\ref{#1})}

\newcommand
{\sect}[1]{\vspace{20pt}{\LARGE}\noindent
{\bf #1:}}
\newcommand
{\subsect}[1]{\vspace{20pt}\hspace*{10pt}{\Large{$\bullet$}}\mbox{ }
{\bf #1}}
\newcommand
{\subsubsect}[1]{\hspace*{20pt}{\large{$\bullet$}}\mbox{ }
{\bf #1}}

\def\ie{{\it i.e.}}
\def\eg{{\it e.g.}}
\def\cf{{\it c.f.}}
\def\etal{{\it et.al.}}
\def\etc{{\it etc.}}

\def\e{{\mbox{{\bf e}}}}
\def\AA{{\cal A}}
\def\BB{{\cal B}}
\def\CC{{\cal C}}
\def\DD{{\cal D}}
\def\EE{{\cal E}}
\def\FF{{\cal F}}
\def\GG{{\cal G}}
\def\HH{{\cal H}}
\def\II{{\cal I}}
\def\JJ{{\cal J}}
\def\KK{{\cal K}}
\def\LL{{\cal L}}
\def\MM{{\cal M}}
\def\NN{{\cal N}}
\def\OO{{\cal O}}
\def\PP{{\cal P}}
\def\QQ{{\cal Q}}
\def\RR{{\cal R}}
\def\SS{{\cal S}}
\def\TT{{\cal T}}
\def\UU{{\cal U}}
\def\VV{{\cal V}}
\def\WW{{\cal W}}
\def\XX{{\cal X}}
\def\YY{{\cal Y}}
\def\ZZ{{\cal Z}}

\def\sinh{{\rm sinh}}
\def\cosh{{\rm cosh}}
\def\tanh{{\rm tanh}}
\def\sgn{{\rm sgn}}
\def\det{{\rm det}}
\def\trace{{\rm Tr}}
\def\exp{{\rm exp}}
\def\sh{{\rm sh}}
\def\ch{{\rm ch}}

\def\ell{{\it l}}
\def\str{{\it str}}
\def\lp{\ell_{{\rm pl}}}
\def\blp{\overline{\ell}_{{\rm pl}}}
\def\ls{\ell_{{\str}}}
\def\bls{{\bar\ell}_{{\str}}}
\def\bM{{\overline{\rm M}}}
\def\gs{g_\str}
\def\gym{{g_{Y}}}
\def\geff{g_{\rm eff}}
\def\eff{{\rm eff}}
\def\r11{R_{11}}
\def\kel{{2\kappa_{11}^2}}
\def\kten{{2\kappa_{10}^2}}
\def\lpten{{\lp^{(10)}}}
\def\alp{{\alpha '}}
\def\alpe{{{\alpha_e}}}
\def\le{{{l}_e}}
\def\aleff{{\alp_{eff}}}
\def\sqaleff{{\alp_{eff}^2}}
\def\tgs{{\tilde{g}_s}}
\def\talp{{{\tilde{\alpha}}'}}
\def\tlp{{\tilde{\ell}_{{\rm pl}}}}
\def\tr11{{\tilde{R}_{11}}}
\def\wtilde{\widetilde}
\def\what{\widehat}
\def\hlp{{\hat{\ell}_{{\rm pl}}}}
\def\hr11{{\hat{R}_{11}}}
\def\hf{{\textstyle\frac12}}
\def\coeff#1#2{{\textstyle{#1\over#2}}}
\def\CY{Calabi-Yau}
\def\lessapprox{\;{\buildrel{<}\over{\scriptstyle\sim}}\;}
\def\greaterapprox{\;{\buildrel{>}\over{\scriptstyle\sim}}\;}
\def\inbar{\,\vrule height1.5ex width.4pt depth0pt}
\def\IC{\relax\hbox{$\inbar\kern-.3em{\rm C}$}}
\def\IR{\relax{\rm I\kern-.18em R}}
\def\IP{\relax{\rm I\kern-.18em P}}
\def\Z{{\bf Z}}
\def\R{{\bf R}}
\def\One{{1\hskip -3pt {\rm l}}}
\def\sst{\scriptscriptstyle}
\def\osc{{\rm\sst osc}}
\def\lam{\lambda}
\def\lc{{\sst LC}}
\def\pr{{\sst \rm pr}}
\def\cl{{\sst \rm cl}}
\def\D{{\sst D}}
\def\bh{{\sst BH}}
\def\vev#1{\langle#1\rangle}

\begin{titlepage}
\rightline{}

\rightline{NSF-KITP-05-89}
\rightline{hep-th/0511196}

\vskip 2cm
\begin{center}
\Large{{\bf Hairy strings}}
\end{center}

\vskip 2cm
\begin{center}
Vatche Sahakian\footnote{\texttt{sahakian@hmc.edu}}
\end{center}
\vskip 12pt
\centerline{\sl Keck Laboratory}
\centerline{\sl Harvey Mudd College}
\centerline{\sl Claremont CA 91711 USA}

\vskip 2cm

\begin{abstract}
Zero modes of the worldsheet spinors of a closed string can source higher order
moments of the bulk supergravity fields. In this work, we analyze
various configurations of closed strings focusing on the imprints
of the quantized spinor vevs onto the tails of bulk fields. We identify 
supersymmetric arrangements for which all multipole charges vanish; while for others, we find that one is left with NSNS and RR dipole and 
quadrupole moments. Our analysis is exhaustive with respect to all the bosonic fields of the bulk and to all higher order moments.
We comment on the relevance of these
results to entropy computations of hairy black holes of a single charge or more, and to open/closed string duality.
\end{abstract}

\end{titlepage}
\newpage
\setcounter{page}{1}

\section{Introduction}
\label{intro}

Recently, much interest has focused on black holes with hair that arise in more than four spacetime dimensions~\cite{Emparan:2001wn}-\cite{Saxena:2005uk}. These new supergravity
solutions seem to teach us a great deal about general relativity in higher
dimensions, as well as string theory. Hairy black holes
may be non-singular and horizon-less; and are often in one to one correspondence with
the microscopic states of the corresponding singular black hole of finite horizon area.
Mathur has suggested~\cite{Mathur:2005zp,Mathur:2005ai} that one is to think of the new smooth solutions
as replacing the old singular black hole geometries in computing certain physical observables (see for example~\cite{Lunin:2001dt}-\cite{Bena:2004wt}). The traditional singular solution is to perhaps correspond to some yet ill-understood average 
geometry of the new smooth ones. In this picture, the horizon radius is mapped onto the size of
a "fuzz" ball beyond which the hairy solutions become indistinguishable from each other and from the original hole. This exciting proposal has been substantiated with various circumstantial and direct evidence. Furthermore, in a class of such black objects where controlled computations may be performed, dipole charges carried by the hole appear to play an important role in correctly accounting for the proper thermodynamics.

One little explored mechanism of correcting and potentially smoothing out traditional black hole solutions involves the veving of the
zero modes of fermions living on the worldvolume of branes or strings~\cite{Plefka:1997xq,Millar:2000ib}. 
Such a treatment would be a semiclassical one where
the leading effect of the quantization of the spinors is taken into account through their back-reaction onto the geometry. The typical length scale for such corrections is set by the worldvolume theory, \ie\ the brane or string tension. Furthermore, fermion vevs would naturally provide hair and multipole moments to the traditional black hole solutions. 

In this work, we concentrate on the IIB fundamental string as a source of the bulk supergravity fields. This system in 
turn may be related by a chain of dualities to various D-brane configurations. Hence, our task is to determine the effect of the spinor vevs of the
IIB Green-Schwarz string onto the IIB supergravity fields. We will approach the
problem from a general framework, gradually narrowing onto a class of doubly charged BPS string configurations. Our starting point is the action given
in~\cite{Sahakian:2004gy}, involving the worldsheet couplings to the supergravity fields
in the light-cone frame. From these couplings, one is able to fix the multipole 
charges of the worldsheet as measured very far away from the worldsheet. 
We consider in turn: a one dimensional beam of massless closed string states, wrapped strings, and wrapped strings with wave profiles along the worldsheet.
In each case, we analyze the effect of the spinor vevs on the asymptotic geometry in preparation to pinning down boundary conditions for new hairy supergravity solutions (for a recent work in this direction, see~\cite{Taylor:2005db}).
This procedure is straightforward, yet leads to certain subtleties; and we find that the final results are indeed very instructive.

In Section 2, we set up the general formalism and present expressions for the
moments in cases where the bulk spacetime is endowed with a set of isometries. In Section 3, we consider  certain worldsheet configurations, and write
the multipole moments for such states explicitely. In Section 4, we describe the
process of veving the spinors and present a concrete example. Section 5 summarizes and reflects on  the results. In the Appendices, we collect
the technical details: general conventions in Appendix A, the worldsheet
couplings in Appendix B, and the map between worldsheet vevs and the string ground states in Appendix C.

\section{Worldsheet couplings and multipole moments}

We consider a class of configurations of IIB closed strings arranged such that the spacetime about the strings is static, and has a spatial isometry along one of the nine space directions. We
work in the light-cone gauge, where the light-cone direction is along the aforementioned isometry of the space - which we label by $x^1$ in the rest of the paper. And our state is to carry $p^+$ units of light-cone momentum. We will interchangeably work in 
scenarios where the light-cone boost direction is a circle of radius $R_s$, or of infinite extent. We will
also refer to the subspace spanned by $x^2,\cdots\,x^9$ as {\em transverse}. Hence, the light-cone is defined by $x^\pm\equiv (x^0\pm x^1)/2$.

Under these circumstances, the couplings of the light-cone Green-Schwarz string to the bulk supergravity fields were worked out in~\cite{Sahakian:2004gy} to all orders in the worldsheet spinors -  
except for couplings to the bulk fermions:
the dilatino and the gravitino. For the setup described in the previous paragraph, 
we may use the results of~\cite{Sahakian:2004gy} to compute the multipole moments of the bulk bosonic fields
far away from the strings. At first sight, one simply reads off from the worldsheet action the ADM
mass in the tail of the metric at infinity, and the NSNS charge of the strings in the tail of the B-field.
We will however consider going beyond this classical treatment: the worldsheet involves 
fermionic degrees of freedom which, for the configurations in question, will have zero modes.
Semiclassically, one sees this effect through the `veving' of fermion bilinears on the worldsheet. These
vevs, as we shall see, can source multipole moments for all of the bulk supergravity fields.  

The action describing IIB supergravity sourced by a closed fundamental string has the structural
form
\bb\label{combinedeq}
\SS=\int d^{10} x\  \LL+\int d^8 x\int d^2\sigma\, \delta^8(x-x_0) \, \LL_s[x_0]\rightarrow \int d^{10} x\  \LL+\int d^{10} x\, \LL_M\ ,
\ee
where $\LL$ is the supergravity Lagrangian, given by~\pref{sugraeq} of Appendix B to linear order in the fields, and $\LL_s$ is the worldsheet Lagrangian of the IIB string given by~\pref{WSeq} in its full glorious non-linear form~\cite{Sahakian:2004gy}. 
We will be interested in configurations of strings confined to the $x^1$ subspace, and our strings are
to be located at some fixed point $x_0^a$ with $a=2,\ldots , 9$, where the $x_0^a$'s are constants on the
worldsheet\footnote{In certain setups that we will be considering, the structural form of
the action given in~\pref{combinedeq} will be reached after a certain averaging prescription. More on this later.}. 

The last arrow in~\pref{combinedeq} involves fixing the light-cone gauge by choosing $x^0(\sigma)$ and $x^1(\sigma)$ using
worldsheet reparameterization invariance. This will allow us to write
\bb\label{jaceq}
d^2\sigma=J\ dx^+ dx^-
\ee
with the Jacobian $J$ treated as a constant on the worldsheet for the cases of interest. Hence, this factor is to be included in $\LL_M$.

The task is then to linearize $\LL_M$ in the supergravity fields in the weak
field approximation regime, and write the corresponding equations of motion for the bulk fields. For the metric $g_{mn}$, one gets
\bb\label{graveq}
(e^a)^n\del^2 \bar{g}_{nm}=-2\kappa^2 \frac{\delta \LL_M}{\delta (e_a)^m}
\ee
where $g_{mn}=\bar{g}_{mn}-\eta_{mn} \bar{g}/8$ with $\bar{g}=\bar{g}_m^m$, and $(e_a)^m$
is the vielbein. Note that one needs to carefully vary the action with respect to the vielbein since one has spacetime
fermions on the Green-Schwarz worldsheet that couple to the spacetime connection. We also have chosen the standard gauge $\del^m \bar{g}_{mn}=0$. 
And $\del^2$ involves the eight transverse directions only.
The right hand side of~\pref{graveq} is straightforward yet somewhat subtle to compute. Our final results will
be collected below.
For the dilaton $\phi$ and axion $\chi$,
the procedure is more straightforward; one gets 
\bb
g_s^2 \del^2 \chi=2\kappa^2 \frac{\delta \LL_M}{\delta \chi}\ ;
\ee
\bb
\del^2\phi=2\kappa^2  \frac{\delta \LL_M}{\delta \phi}\ .
\ee
While for the 2-form gauge field $A_{mn}$, one gets
\bb
\del^2 A^{mn}=6 \kappa^2  \frac{\delta \LL_M}{\delta \bar{A}^{mn}}
\ee
with the gauge condition $\del^m A_{mn}=0$. 
$A_{mn}$ is complex and involves both NSNS and 
RR pieces $A^{(1)}_{mn}$ and $A^{(2)}_{mn}$ respectively
\bb
A_{mn}=\frac{1}{2\sqrt{g_s}}A^{(1)}_{mn}+i \frac{\sqrt{g_s}}{2} A^{(2)}_{mn}\ ,
\ee
as described in Appendix B\footnote{Note that these simplified relations are valid in
the linearized approximation scheme that we need.}.
While the real 4-form gauge field obeys
\bb
\del^2 A^{mnpq}=96 \kappa^2  \frac{\delta \LL_M}{\delta \bar{A}^{mnpq}}
\ee
with $\del^m A_{mnpq}=0$. 

Since we are working at the linearized level in the supergravity fields, at this stage all derivatives may be treated as
arising from a flat connection; and there is no need to distinguish between tangent space indices
$a,b,c,\cdots$ and spacetime indices $m,n,p,\cdots$. We will hence revert to tangent space indices for the rest of the paper. Furthermore, $a,b,c,\cdots$ will always label the eight transverse
directions.

Using equations~\pref{WSeq}-\pref{lasteq}, one finds:

\begin{itemize}
\item  A dipole moment of the metric of the form
\bbb\nonumber
g^{ab}&=&-J\,\imag \,{\pi }^2\,{\alpha'}^3\,{\sqrt{{g_s}}}\,
  {{{V_i}}^+}\,\left[  - 8\,{\sqrt{-h}}\,h^{{ij}}\,
     {\bar{\theta}}
      {\sigma }^{-{c(a}}\theta\,{{V_j}}^{b)}
    +8\,{\epsilon }^{{ij}}\,
     \theta{\sigma }^{-{c(a}}
      \theta\,{{V_j}}^{b)} 
      \re. \\ &+&\lk. 
      \eta^{ab}\, {{V_{j\, d}}}\, \left(
       {\sqrt{-h}}\,h^{{ij}}\,
        {\bar{\theta}}
         {\sigma }^{-{cd}}\theta- {\epsilon }^{{ij}}\,
          \theta{\sigma }^{-{cd}}
           \theta    \right) \,
    \right] \,{{\partial }_c}\lk(\frac{1}{r^6}\re)+\mbox{c.c}\ ;\label{metricdipole}
\eee
where
\bb
V_i^a=\del_i x^a\ \ \ ,\ \ \ V_i^+=\del_i x^+\ ,
\ee
with $i,j,\cdots$ labeling the worldsheet directions $\sigma^0$ and $\sigma^1$. The $\sigma^{-ab\cdots}$s
are gamma matrices and the $\theta$'s are the Green-Schwarz spinors, both defined
in Appendix A.  In this and similar subsequent expressions, one is to think of the spinors 
$\theta$ as `veved'; \ie\ one is to take the expectation value of~\pref{metricdipole}, $\langle g^{ab}\rangle$, in the
state space of the spinor zero modes. In Section 4, we will look at explicit realizations of
such vevs. Note that $J$ is the Jacobian defined in~\pref{jaceq}.
One also finds a quadrupole moment given by
\bbb\nonumber
g^{ab}&=&J \frac{{\pi }^2}{24}\,{\alpha'}^3\,
    {\sqrt{{g_s}}}\,{\sqrt{-h}}\,h^{{ij}}\,{{{V_i}}^+}\,{{{V_j}}^+}\,
    \left[ 4\,{\bar{\theta}}
        {\sigma }^{-{c(a}}\theta\,
       {\bar{\theta}}
        {\sigma }^{b)-{d}}\theta 
      \re. \\ &+&\lk.         
         4\,\eta^{d(b}\,{\bar{\theta}}
        {\sigma }^{a)-{e}}\theta\,
       {\bar{\theta}}
        {\sigma }_{\ \,e}^{-{\ c}}\theta 
        +\eta^{ab}\, {\bar{\theta}}
        {\sigma }^{-{ec}}\theta\,
       {\bar{\theta}}
        {\sigma }^{{- d}}_{\ \ \ e}\theta\right] \,
    {{\partial }_c}{{\partial }_d}\lk(\frac{1}{r^6}\re)+\mbox{c.c}\ .
\eee
No higher multipole fields arise in the light-cone gauge from the spinor vevs. Note that the
latin indices label only the eight transverse directions\footnote{However, in certain cases, one
may also read-off the $g^{0a}$, $g^{1a}$, and $g^{01}$ components from these expressions. The subtlety arises because of certain assumptions about the structure of the background fields made in~\cite{} in
attaining equation~\pref{WSeq}. Particularly, one must be careful in this formalism while trying to extract the spins of the closed
string states from moments of the metric.}.

\item A dipole moment encoded in the dilaton field
\bb
\phi=4\,J\,\imag \,{\pi }^2\,{\alpha'}^3\,{\sqrt{{g_s}}}\,
   {{{V_i}}^+}\,{\epsilon }^{{ij}}\,\,{{V_{j\, a}}}
      {\bar{\theta}}
        {\sigma }^{-{ab}}{\bar{\theta}} 
    \,{{\partial }_b}\lk(\frac{1}{r^6}\re)+\mbox{c.c}\ ;\label{dilatondipole}
\ee
and in the axion field
\bb
\chi={-4\,J\,{\pi }^2}\,\frac{{\alpha'}^3}{{\sqrt{{g_s}}}}\,
   {{{V_i}}^+}\, \left[ {\sqrt{-h}}\,h^{{ij}}\,{{V_j}}^b\,
       {\bar{\theta}}\sigma^-\theta
       -{\epsilon }^{{ij}}\,{{V_{j\, a}}}\,
         {\bar{\theta}}
          {\sigma }^{-{ab}}{\bar{\theta}}
      \right]\,{{\partial }_b}\lk(\frac{1}{r^6}\re)+\mbox{c.c}\ ;\label{axiondipole}
\ee
Again, no higher moments arise from the spinor vevs.

\item Magnetic and electric dipole moments of the 2-form gauge field
\bbb\nonumber
{A}^{ab}&=&-12\,J\,\imag \,{\pi }^2\,{\alpha' }^3\,{\sqrt{{g_s}}}\,
  {{{V_i}}^+}\, \lk[
    {\sqrt{-h}}\,h^{{ij}}\,
     \left( -2\,{{\theta}}
         {\sigma }^{-{c[a}}{{\theta}}\,{{V_j}}^{b]} - 
       {{\theta}}
         {\sigma }^{-{ab}}{{\theta}}\,{{V_j}}^c \right) \re.
        \\ &+&\left.
         {\epsilon }^{{ij}}\,
     \left(  2\, {\bar{\theta}}
         {\sigma }^{-{c[a}}\theta\,{{V_j}}^{b]} + 
       {\bar{\theta}}
         {\sigma }^{-{ab}}\theta\,{{V_j}}^c \right)  
    \right]\,{{\partial }_c}\lk(\frac{1}{r^6}\re)+\mbox{c.c}\ ;\label{twoformdipole1}
\eee
\bbb
{A}^{+-}&=&-48\,J\,\imag \,{\pi }^2\,{\alpha' }^3\,{\sqrt{{g_s}}} \,{{{V_i}}^+}\,
  \left[ {-\sqrt{-h}}\,h^{{ij}}\,{{V_j}}^a\,
     {{\theta}}
      {\sigma }^{-{ab}}{{\theta}}
      \re. \nonumber \\ &-& \lk.
    {\epsilon }^{{ij}}\,
     {{V_j}}^a \, {\bar{\theta}}
         {\sigma }^{-{ab}}\theta
        \right]\,{{\partial }_b}\lk(\frac{1}{r^6}\re)+\mbox{c.c}\ ;\label{twoformdipole2}
\eee
This time, we see explicitly components in the light-cone directions. And one also finds some  
electric quadrupole moments in this field
\bb
A^{ab}=0
\ee
\bb
A^{+-}=4\,J\,{\pi }^2\,{\alpha' }^3\,
 {\sqrt{{g_s}}}\,{\sqrt{-h}}\,h^{{ij}}\,{{{V_i}}^+}\,{{{V_j}}^+}\,
   \theta{\sigma }^{-{cb}}\theta\,
  {\bar{\theta}}{\sigma }^{-{a}}_{\ \ \ c}
   \theta\,{{\partial }_a}{{\partial }_b}\lk(\frac{1}{r^6}\re)+\mbox{c.c}\ ;
\ee
Once again, no higher moments arise.

\item Finally, a dipole moment of the 4-form gauge field
\bb
A^{+-ab}=192\,J\,{\pi }^2\,{\alpha' }^3\,
  {\sqrt{{g_s}}}\,{\epsilon }^{{ij}}\,{{{V_i}}^+}\,\left[ 2\,\theta
      {\sigma }^{-{c[b}}\theta\,{{V_j}}^{a]} - 
    \theta{\sigma }^{-{ab}}
      \theta\,{{V_j}}^c \right]\,{{\partial }_c}\lk(\frac{1}{r^6}\re)+\mbox{c.c}\ ;\label{fourformdipole}
\ee
with all other components being zero.
\end{itemize}

These multipole charges are relaying information about the fermionic condensates
to infinity. We want to think of these expressions throughout as sandwiched between states of the quantized zero modes of the spinors; \ie\ for the axion
\bb
\chi\rightarrow \Bra{Z_1} \chi \Ket{Z_2}
\ee
where $\Ket{Z_1}$ and $\Ket{Z_2}$ would be one of the 256 ground states of the closed 
string arising from the Clifford algebra of the zero modes.
Additional information may be hidden in the moments of the dilatino and gravitino bulk fields. At the linearized level,
these additional couplings on the worldsheet do not affect our results. However,
a complete picture of all the information about the closed string state as
seen from far away requires knowledge of these additional couplings.

Our next task will be to explore particular interesting arrangements of closed strings
that fit the general requirements we outlined at the beginning of this section.

\section{Examples}

In this section, we look at four different configurations of closed strings whose
spacetime imprints share the
general features required of the bulk fields in the previous section. Focusing on specific examples
allows us to write more explicit formulae for the moments in certain common scenarios. This will also
help us in acquiring some basic physical intuition about the role of these moments in describing
closed string states.

For the different setups we consider, we will need to fix the worldsheet reparametrization and Weyl scale invariance by fixing
the worldsheet metric as in $h_{00}=-h_{11}=1$ and $h_{01}=0$~\footnote{Note that in our conventions the signature of the metric is mostly negative.}, along with the coordinates $x^\pm(\sigma)$ 
while
assuring that the Virasoro constraints are satisfied
\bb\label{virasoroeq}
\lk(\dot{x}\pm x'\re)^2=0\ .
\ee
We will then need to determine the Jacobian~\pref{jaceq}. Finally, we will put things together carefully so as not to violate the assumed isometries of the bulk spacetime. Note that one is to arrange the worldsheet configuration as a source {\em in
flat space}; the back-reaction of the spacetime onto the source is subleading to
this computation. Hence, the Virasoro constraint above does not include the spinors
as these arise in terms coupled to field strengths and the connection.

\subsection{Standard ground state}

Our first example is the simplest possible setup. We consider `point'-like closed string states
\ie\ the 256 ground states of the closed string; we pick one species, and
consider a uniform density beam of these massless particles projected along the $x^1$ direction. For
each particle, one has
\bb
x^+=p^+\frac{\alpha'}{\sqrt{g_s}}\sigma^0\ \ \ ,\ \ \ x^-=0\ ,
\ee
with $x^a=\mbox{constant}$.
This setup requires us to consider the Green's function in 9 space dimensions, instead of 8,
as given by~\pref{greeneq}. Hence, the measure of the worldsheet action takes the form
\bb
\int d^8 x\,dx^1\, \delta^8(x) \sum_n \delta(x^1-n \epsilon) \int d^2\sigma\ ,
\ee
where we have laid the particles in the beam with a small spacing $\epsilon$ along $x^1$.
In a semiclassical treatment, the ground state is not point like but has a small size, say $\epsilon$
(of order the string scale). Hence, we write
\bb\label{fac1eq}
\int d\sigma\rightarrow \epsilon\ .
\ee
Integrating over $x^1$, we would need to deal with the new Green's function
\bb\label{fac2eq}
\sum_n \frac{1}{((x^1-n\epsilon)^2+x^a x^a )^{7/2}}\rightarrow \frac{16}{15} \frac{G_8}{\epsilon}\ ,
\ee
where $G_8$ is the eight dimensional Green's function $1/r^6$, and the arrow implies
turning the sum into an integral in the limit $\epsilon\rightarrow 0$. One also needs a factor 
of $7\Omega_8/6\Omega_7=15/16$. Putting things together,
the two $\epsilon$ factors from~\pref{fac1eq} and~\pref{fac2eq} cancel and $\LL_M$ in~\pref{combinedeq} acquires an
additional factor of
\bb
J=\frac{1}{p^+} \frac{g_s^{3/4}}{{\alpha'}^{3/2}}\ ;
\ee
Otherwise, one may now make use of the eight dimensional Green's function in moment computations.

In this scenario, one finds the following simplified expressions for the moments from~\pref{metricdipole}-\pref{fourformdipole}:
\begin{itemize}
\item The dipole moment of the metric is
\bb
g_{ab}=0\ ;
\ee
the quadrupole moment becomes
\bbb\nonumber
g^{ab}&=&\frac{1}{96}\,{\pi }^2\,{\left(p^+\frac{\sqrt{\alpha'}}{g_s^{1/4}}\right)}\,{\alpha' }^3\,{\sqrt{{g_s}}}\,
    \left[ 16\,{\bar{\theta}}
        {\sigma }^{-{c(a}}\theta\,
       {\bar{\theta}}
        {\sigma }^{b)-{d}}\theta 
                      \re. \\ &-&\lk. 
     16\,{{\eta }^{d(a}}\,{\bar{\theta}}
        {\sigma }^{b)-{e}}\theta\,
       {\bar{\theta}}
        {\sigma }^{{-c}}_{\ \ \ e}\theta 
        -4\,{{\eta }^{{ab}}}\,{\bar{\theta}}
        {\sigma }^{-{ec}}\theta\,
       {\bar{\theta}}
        {\sigma }^{{-d}}_{\ \ \ e}\theta
      \right] \,{{\partial }_c}{{\partial }_d}\lk(\frac{1}{r^6}\re)+\mbox{c.c}\ ;
\eee

\item No moments for the dilaton
\bb
\phi=0\ ;
\ee
and none for the axion field
\bb
\chi=0\ .
\ee

\item No dipole moment for the 2-form
\bb
A_{ab}=0\ ;
\ee
but quadrupole moments may exist
\bb
A^{ab}=2\,{\pi }^2\,{\left(p^+\frac{\sqrt{\alpha'}}{g_s^{1/4}}\right)}\,{\alpha' }^3\,{\sqrt{{g_s}}}\,
    \theta{\sigma }^{-{ad}}
     \theta\,{\bar{\theta}}
     {\sigma }^{-{bc}}\theta\,
    {{\partial }_c}{{\partial }_d}\lk(\frac{1}{r^6}\re)+\mbox{c.c}\ ;
\ee
\bb
A^{+-}=-4\,{\pi }^2\,{\left(p^+\frac{\sqrt{\alpha'}}{g_s^{1/4}}\right)}\,{\alpha' }^3\,{\sqrt{{g_s}}}\,
    \theta{\sigma }^{-{bc}}
     \theta\,{\bar{\theta}}
     {\sigma }^{-{ac}}\theta\,
    {{\partial }_a}{{\partial }_b}\lk(\frac{1}{r^6}\re)+\mbox{c.c}\ ;
\ee

\item And finally no moments for the 4-form
\bb
A^{+-ab}=0\ .
\ee
\end{itemize}
Note that these equations are written in the Einstein frame: a scaling $\alpha'\rightarrow \alpha'\sqrt{g_s}$ yields to the canonical factor $g_s^2$ of the string frame (after rescaling $\chi$ as well). Hence, we see from these expressions that we have the possibility of non-zero RR electric and magnetic quadrupole moments in the 2-form gauge field.

\subsection{Wrapped ground state}

Our next example is that of a closed string wrapping the $x^1$ direction, which is
also the longitudinal direction of the light-cone gauge. Hence, we might want to think 
of the $x^1$ direction as being compact. Since the Green's function in question does not probe
this direction, compactifying it is harmless. Due to the Virasoro constraints~\pref{virasoroeq},
one needs
\bb
x^+=p^+\lk(\sigma^0+\sigma^1\re)\ \ \ ,\ \ \ x^-=p^+\lk(\sigma^0-\sigma^1\re)\ \ \ \mbox{or}\ \ \ 0\ ,
\ee
with $x^a=0$.
One then easily finds that all moments~\pref{metricdipole}-\pref{fourformdipole}  vanish! A wrapped BPS string does not carry any multipole moments.

\subsection{Wrapped ground state with a parallel wave}

A more interesting setup would be to consider the wrapped state of the previous section
\bb
x^1=w R_s \sigma^1\ ;
\ee
where $w$ is the wrapping number and $R_s$ is the radius of compactification;
but to add along the string a wave of mode number $n$
\bb\label{coseq}
x^2=A\cos\lk[ n(\sigma^0+\sigma^1)\re]\ ,
\ee
{\em parallel} to the wrapping. To assure that the spacetime retains the needed
isometries, we would need that the amplitude of the wave $A$
is small enough $A\ll R_s$, and that observations are made over time scales large compared to the period of oscillation; alternatively, we would increase $n\gg 1$. This would allow us to average
over the wave profile. We may even 
consider a more general profile
\bb
x^2=f(\sigma^0+\sigma^1)\ ,
\ee
as long as $f$ is such that the aforementioned assumptions may still be made.
Note that such states with unidirectional momentum modes are 1/2 BPS.
We first solve
the Virasoro constraint~\pref{virasoroeq}
\bb
x^0=w R_s \sigma^0 +F(\sigma^0+\sigma^1)
\ee
where $F'$ can easily be found
\bb
F'=-w R_s+ \sqrt{w^2 R_s^2 +{f'}^2}\ .
\ee
We then have
\bb\label{zeroquadeq}
V^+_0=V^+_1=\frac{1}{2}\lk(F'+w R_s\re)\ .
\ee
Even before averaging over the profile $f$, one can easily see that all quadrupole moments vanish since $V^+_i V^{+i}=0$! This arises from the fact that the momentum modes run parallel to the
wrapping. Hence,
a 1/2 BPS state consisting of a wrapped closed string with a plane wave moving parallel to the
wrapping direction gives no quadrupole moments.

As for dipole moments, we see from~\pref{metricdipole}, \pref{dilatondipole}, \pref{axiondipole},
\pref{twoformdipole1}, \pref{twoformdipole2}, and~\pref{fourformdipole}
that these all involve the factors $h^{ij} V^a_i V^+_j$ or $\varepsilon^{ij} V^a_i V^+_j$. From
~\pref{zeroquadeq}, we immediately see that all dipole moments vanish as well. In reaching this result,
the fact that the wave on the string is moving parallel to the winding direction is crucial. Hence, this
system does not carry any dipole charges either.

\subsection{Wrapped ground state with an antiparallel wave}

The obvious next step is to put the wave on the wrapped string of the previous section
in the opposite direction to the wrapping
\bb\label{cos2eq}
x^2=A\cos \lk[n(\sigma^0-\sigma^1)\re]\ .
\ee
Or in general
\bb
x^2=f(\sigma^0-\sigma^1)
\ee
We again solve the Virasoro constraint with
\bb
x^0=w R_s \sigma^0 +F(\sigma^0-\sigma^1)
\ee
for $F$ again given by
\bb
F'=-w R_s+ \sqrt{w^2 R_s^2 +{f'}^2}\ .
\ee
We define the shorthand
\bb
K\equiv \frac{1}{2} \sqrt{w^2 R_s^2 +{f'}^2}\ ; 
\ee
We then find, unlike in~\pref{zeroquadeq}
\bb
V^+_0+V^+_1=w R_s\ ,
\ee
for any profile $f$, with
\bb
V^+_0=K\ .
\ee
The Jacobian of the measure is
\bb
dt\,dx=2 w R_s K\ d^2\sigma=2 {d^2\sigma}/{J}\ .
\ee
We now need to insert these expressions into~\pref{metricdipole}-\pref{fourformdipole}. 
We focus next on the needed averaging process. We encounter two types of
factors to average over
\bb
\left\langle\frac{f'}{2 K}\right\rangle\equiv D
\ee
in dipole moments, and
\bb
\left\langle\frac{2 K-w R_s}{2 K} \right\rangle\equiv Q
\ee
in quadrupole moments.

The averages are to be taken over timescales much larger than
the oscillation period of the profile, which is set by the string scale. From
the viewpoint of the bulk spacetime, superstringy timescales of observation are indeed natural and needed\footnote{For short wavelengths or large mode numbers, we may write for example
$\langle F \rangle\equiv \frac{1}{2\pi} \int_0^{2\pi} d\sigma\,F\ .$}.

We look at two cases:
\begin{enumerate}
\item For $w R_s\gg f'$, on has
\bb\label{case1eq}
D\simeq -\frac{1}{2} \frac{\langle {f'}^3\rangle}{(w R_s)^3}\ll 1\ ,
\ee
and
\bb
Q\simeq \frac{\langle {f'}^2\rangle}{2 w^2 R_s^2}\ll 1\ ;
\ee
where we have used the fact that $f$ must be periodic $f(2\pi)=f(0)$. These expressions are
finite but parametrically small with the average amplitude of the profile.
\item For $w R_s\ll f'$, one gets
\bb\label{case2eq}
D\simeq\langle \sgn(f') \rangle<1\ ,
\ee
and
\bb
Q\simeq 1\ .
\ee
which are finite and may be of order 1.
\end{enumerate}

Both cases are to be arranged so that the perturbations along the strings are of negligible amplitude compared
to the scale set by $R_s$. For case one, one needs to have low mode numbers; for case two,
one needs large mode numbers (and the averaging process gets favored even more by the supergravity side). 
In either case, one is arranging for profiles that appear homogeneous over long enough timescales, yet
give the strings a `thickness' in the $x^2$ direction. Furthermore, in both cases, all profiles $f$ that are even functions yield zero dipole moments.

We then find the following moments:
\begin{itemize}
\item A dipole moment for the metric
\bbb\nonumber
g^{ab}&=&{\imag{{\pi }^2}} \,D\,{\alpha' }^3\,
    {\sqrt{{g_s}}}\,\left[ -16\,{{\eta }^{2(b}}\,
       \theta{\sigma }^{a)-{c}}
        \theta - 
      16\,{{\eta }^{2(b}}\,{\bar{\theta}}
        {\sigma }^{a)-{c}}\theta +
       2\,{{\eta }^{{ab}}}\,\theta{\sigma }^{-{c2}}
        \theta 
              \re. \\ &+&\lk. 
        2\,{{\eta }^{{ab}}} \,{\bar{\theta}}
        {\sigma }^{-{c2}}\theta 
        \right] \,{{\partial }_c}\lk(\frac{1}{r^6}\re)+\mbox{c.c}\ ;
 \eee
And a non-zero quadrupole moment as well
\bbb\nonumber
g^{ab}&=&\frac{{{\pi }^2}}{48}\,
   Q\,{\alpha' }^3\,{\sqrt{{g_s}}} \,
    \left[  
      16\,{\bar{\theta}}
        {\sigma }^{-{c(a}}\theta\,
       {\bar{\theta}}
        {\sigma }^{b)-{d}}\theta 
              \re. \\ &-&\lk. 
        4\,{{\eta }^{{ab}}}\,{\bar{\theta}}
        {\sigma }^{-{ce}}\theta\,
       {\bar{\theta}}
        {\sigma }^{-{d}}_{\ \ \ e}\theta 
       -16\,{{\eta }^{{d(b}}}\, {\bar{\theta}}
        {\sigma }^{a)-{e}}\theta\,
       {\bar{\theta}}
        {\sigma }^{-{c}}_{\ \ \ e}\theta 
              \re. \\ &+&\lk. 
      8\,{{\eta }^{{cd}}}\,{\bar{\theta}}
        {\sigma }^{-{ea}}\theta\,
       {\bar{\theta}}
        {\sigma }^{-{b}}_{\ \ \ e}\theta +
       {{\eta }^{{ab}}}\,{{\eta }^{{cd}}}\,{{\bar{\theta}}
          {\sigma }^{-{ef}}\theta}\,{{\bar{\theta}}
          {\sigma }^{-}_{{\ \ ef}}\theta} 
       \right] \,
    {{\partial }_c}{{\partial }_d}\lk(\frac{1}{r^6}\re)+\mbox{c.c}\ ;
\eee

\item Dipole moments for the dilaton and axion
\bb
\phi={ 8 \imag{\pi }^2}D\,{\alpha' }^3\,
    {\sqrt{{g_s}}}\,\left[ -{\bar{\theta}}
       {\sigma }^{-{b2}}{\bar{\theta}} + 
       {{\eta }^{{b2}}\,{\bar{\theta}}\sigma^-\theta} \right] \,{{\partial }_b}\lk(\frac{1}{r^6}\re)+\mbox{c.c}\ ;
\ee
\bb
\chi={8\,{\pi }^2}\,D\,\frac{{\alpha' }^3}{{\sqrt{{g_s}}}}\,
    \left[ -{\bar{\theta}}
       {\sigma }^{-{b2}}{\bar{\theta}} + 
       {{\eta }^{{b2}}\,{\bar{\theta}}\sigma^-\theta} \right] \,{{\partial }_b}\lk(\frac{1}{r^6}\re)+\mbox{c.c}\ ;
\ee

\item Dipole moments for the 2-form
 \bbb\nonumber
{A}^{ab}&=&{-24\,\imag}\,{\pi }^2 \,D\,{\alpha' }^3\,
    {\sqrt{{g_s}}}\,\left[ 
     2\,{{\eta }^{{2[b}}}\, {\bar{\theta}}
        {\sigma }^{a]-{c}}\theta +
       2\,{{\eta }^{{2[b}}} \,{{\theta}}
        {\sigma }^{a]-{c}}{{\theta}}
             \re. \\ &+&\lk. 
       {{\eta }^{{c2}}}\,{\bar{\theta}}
        {\sigma }^{-{ab}}\theta +
       {{\eta }^{{c2}}\,{{\theta}}
        {\sigma }^{-{ab}}{{\theta}}} \right] \,{{\partial }_c}\lk(\frac{1}{r^6}\re)+\mbox{c.c}\ ;
\eee
\bb
{A}^{+-}={-96\,\imag\,{\pi }^2}\,D\,{\alpha' }^3\,
    {\sqrt{{g_s}}}\,\left[ {\bar{\theta}}
       {\sigma }^{-{b2}}\theta - 
      {{\theta}}
       {\sigma }^{-{b2}}{{\theta}} - 
      {{\eta }^{{b2}}}\,{\bar{\theta}}\sigma^-\theta
        \right] \,{{\partial }_b}\lk(\frac{1}{r^6}\re)+\mbox{c.c}\ ;
\ee
And a non-zero electric quadrupole moment as well
\bb
A^{ab}=0\ ;
\ee
\bb
A^{+-}={-8}\,{\pi }^2 Q\,{\alpha' }^3\,
    {\sqrt{{g_s}}}\,
    \,\theta{\sigma }^{-{bc}}
     \theta\,{\bar{\theta}}
     {\sigma }^{-{a}}_{\ \ \ c}\theta\,{{\partial }_a}{{\partial }_b}\lk(\frac{1}{r^6}\re)+\mbox{c.c}\ ;
\ee

\item Finally, an electric dipole moment of the 4-form
\bb
A^{+-ab}={-384}\,{\pi }^2\,D\,{\alpha' }^3\,{\sqrt{{g_s}}}\,
    \left( -2\,{{\eta }^{{2[a}}}\,\theta
        {\sigma }^{b]-{c}}\theta +
       {{\eta }^{{c2}}}\,\theta{\sigma }^{-{ab}}
        \theta \right) \,
    {{\partial }_c}\lk(\frac{1}{r^6}\re)+\mbox{c.c}\ ;
\ee
\end{itemize}

We see that one may have a myriad of multipole charges for wrapped strings with antiparallel modes
on the worldsheet - in contrast to the case where the modes are running parallel to the wrapping.

\section{Polarization states}

Next, we would like to consider as a concrete example an explicit condensate
of the spinors on the worldsheet. To do this, one needs to describe the state space of
zero modes of the spinors in the standard way~\cite{}. We first
decompose our Weyl spinors $\theta$ into Majorana-Weyl spinors $\theta_1$ and $\theta_2$
\bb
\theta\equiv\theta_1+i \theta_2\ \ \ ,\ \ \ \bar{\theta}\equiv\theta_1-i \theta_2\ .
\ee
We define states $\Ket{A}$ and $\Ket{\alpha}$ where $A=2,\cdots,9$ describes the
polarizations of the vector
state, and $\alpha=1,\cdots,16$ labels the polarizations of the spinor; the latter
satisfying $(\sigma^+)^{\alpha}_\beta \Ket{\alpha}=0$, leaving 8 polarizations. 
Requiring that the rotation operator $\overbar{\theta}\sigma^{-ab}\theta$ acts on these states properly, we
write
\bb
\theta_1^\alpha \Ket{A}_1=\frac{\sqrt{\Gamma}}{2} \lk(\sigma^A\re)^{\alpha\beta}\Ket{\beta}_1\ .
\ee
\bb
\theta_1^\alpha \Ket{\beta}_1=\frac{\sqrt{\Gamma}}{2} \lk(\sigma^{+}_{\,\,\,\,a}\re)^\alpha_{\,\beta}\Ket{a}_1\ .
\ee
where $\Gamma$ is fixed by the quantization condition $\{\theta^\alpha,(\sigma^-\overbar{\theta})_\beta\}=\Gamma\, \delta^\alpha_\beta$ of the spinors
\bb
\Gamma\equiv \frac{1}{p^+}\ .
\ee

For the closed string, the state space is spanned by a direct product of two copies of such states
\bb
\Ket{A,B}\ \ \ ,\ \ \ \Ket{A,\alpha}\ \ \ ,\ \ \ \Ket{\alpha,A}\ \ \ ,\ \ \ \Ket{\alpha,\beta}\ ,
\ee
yielding the usual 256 states of the IIB supergravity multiplet.
We normalize these as in\footnote{Note that the signature of our metric is mostly negative.}
\bb
\langle A | B \rangle=-\eta^{AB}\ \ \ ,\ \ \ 
\langle \alpha | \beta \rangle=\delta_{\alpha\beta}\ .
\ee
The task then is to take the expectation value of the expressions in the previous sections in
the desired states. This is straightforward but rather tedious - requiring extensive use of Fierz identities.
We have tabulated the needed background information in Appendix C, which includes all possible 
vevs that can arise in any such computation.

As a simple demonstration, focus on the simplest state: a
dilatonic state, described by $\Ket{A,A}$ where $A$ is being summed over. Using the catalogue in
Appendix C, one finds that all moments in~\pref{metricdipole}-\pref{fourformdipole} vanish except two:

\begin{itemize}
\item The axion dipole moment:
\bb
\chi=-\frac{512\,{\pi }^2}{p^+}\,J\,\frac{{\alpha'}^3}{{\sqrt{{g_s}}}}\,{\sqrt{-h}}\,h^{i\,j}\,{{V_i}}^+\,{{V_j}}^a\,{{\del }_a}(r^{-6})
\ee

\item And the quadrupole moment of the metric
\bb
g_{ab}=-\frac{2^{10}\,{\pi }^2}{(p^+)^2}\,J\,{\Gamma }^2\,{\alpha'}^3\,{\sqrt{{g_s}}}\,{\sqrt{-h}}\,h^{i\,j}\,{{V_i}}^+\,{{V_j}}^+\,{{\del }_a}{{\del }_b}(r^{-6})
\ee
\end{itemize}

Depending on the embedding in the target space, one can get non-zero moments in
the axion field or metric. For example, for a wrapped configuration with an antiparallel wave
in the dilaton state polarized in the $x^2$ direction, we would have an RR axion dipole moment given by
\bb\label{chi}
\chi=\frac{3072\pi^2}{p^+} \frac{{\alpha'}^3}{\sqrt{g_s}} D \frac{x^2}{r^8}\ ,
\ee
with $D$ defined in~\pref{case1eq} or~\pref{case2eq}.

\section{Discussion and results}

In this work, we considered a class of IIB string configurations
which have the zero modes of the worldsheet spinors veved through quantization. By focusing on the moments of the bosonic fields, we unravelled 
a map between worldsheet states and the multipole charges as seen at infinity. Our conclusions may be summarized as follows:

\begin{itemize}
\item Spinor vevs may source dipole and quadrupole moments in general but no higher moments.

\item For the beam of ground state particles, and for strings wrapping a cycle of the geometry with or without a parallel wave along it, all moments from spinor vevs vanish.

\item For wrapped strings with waves along it arranged antiparallel to the winding, we find in general non-zero dipole and quadrupole moments of both RR and NSNS types. 

\item We presented a catalogue for computing moments for BPS as well as non-BPS string configurations, provided the bulk spacetime retains certain
isometries.
\end{itemize}

These results fix the asymptotic tails of bulk fields in new geometries that one is to look for and that may exist in one to one correspondence with IIB string states of different zero mode polarizations. Consider for example the case of a singly charged hole. Perhaps the spinor vevs are to react on the geometry so as to generate new
fuzz-ball solutions; yet, the area of the fuzz-ball should be proportional to the log of the degeneracy of the ground state, \ie\ log of 256 for a singly charged hole. This implies
that the area of the `stretched horizon'
is of order the string scale. Hence, the corresponding supergravity solutions would be
expected to break down in the region of most interest. The more interesting case is that of a hole with two charges, such as one corresponding to a IIB string with an antiparallel wave. 
Here, we may expect the existence of new smooth geometries with dipole and quadrupole moments\footnote{In a recent work, Taylor provides explicit solutions for strings with classical bosonic and fermionic worldsheet profiles~\cite{Taylor:2005db}. This work explores only NSNS couplings by working in the worldsheet picture as opposed to the Green-Schwarz formalism we have adopted. The hallmark of spinor condensates appearing as multipole moments in the bulk fields is seen in that work as well.}. However, spinor vevs can also have an imprint onto the tail of the dilatino and gravitino, which we did not consider. Adding the effect  of the couplings of these supergravity fields to the formalism greatly complicates matters, and would be needed to a proper accounting of the entropy. Hence, a complete treatment is deferred to a future work. Another direction of pursuit would involve non-BPS configurations, with both right and left moving waves on the string.

Another interesting aspect of these results has to do with open/close string duality. For example in the case of equation~\pref{chi}, we identified a certain IIB string state that is endowed with D-instaton
dipole moment. One may then ask whether there is a cartoon involving,
say, a D-instaton and an anti-D-instanton that mimics the corresponding IIB string state. Indeed, in the context of Matrix theory~\cite{Banks:1996vh,Dijkgraaf:1997vv}, the fundamental string is realized through the degrees of freedom of wound D-strings. In this spirit, it is easiest to attempt to construct such a picture of the
closed string states with D-instantons only, guided by the required multipole moments of the axion. The phenomena of D-brane polarization \`{a} la Myers~\cite{Myers:1999ps} would generate the expected multipole moments for the other brane charges for free (\ie\ because of T-duality). We hope to report on this in the near future. 

\vspace{0.5in}
{\bf Acknowledgments}

I am grateful to the KITP for hospitality while this paper was completed. This work was supported in part by the Research Corporation Grant No. CC6483 and the NSF Grant No. PHY99-07949. 

\section{Appendices}

\subsection{Appendix A: Spinor conventions}

Our spinors are Weyl but not Majorana. They are complex and have
sixteen components. The associated $16\times 16$ gamma matrices satisfy
\bb
\lk\{\sigma^a,\sigma^b\re\}=2\eta^{ab}\ ,
\ee
with the metric
\bb
\eta_{ab}=diag(+1,-1,-1,...,-1)\ .
\ee
Note that the signature is different from
the standard one in use in modern literature. This is so that
we conform to the equations appearing in~\cite{Howe:1983sr}.
Also, the worldsheet metric $h^{ij}$ has signature $(-,+)$
for space and time, respectively.
Throughout,
the reader may refer to~\cite{Howe:1983sr} to determine more about the
spinorial algebra and identities that we are using. However, 
we make no distinction between $\sigma$ and $\hat{\sigma}$ as defined
in~\cite{Howe:1983sr} as this will be obvious from the context.

We note that
$\sigma^a$, $\sigma^{abcd}$ and $\sigma^{abcde}$ are symmetric; while
$\sigma^{ab}$ and $\sigma^{abc}$ are antisymmetric; and
$\sigma^{abcde}$ is self-dual.
We then have
\bb
\sigma^+\sigma^-+\sigma^-\sigma^+=1\ .
\ee
And complex conjugation is defined so that
\bb
\overline{\sigma^a}=\sigma^a\ .
\ee
Conjugation also implies
\bb
\overline{\theta_1 \theta_2}=\bar{\theta}_2 \bar{\theta}_1\ .
\ee
Finally, antisymmetrization is defined as
\bb
\sigma^{ab}\equiv \sigma^{[a} \sigma^{b]}\ ,
\ee
with a conventional $2!$ hidden by the braces. For more details about the
conventions and the formalism used, the reader is referred to~
\cite{Sahakian:2004gy}.

\subsection{Appendix B: The worldsheet and bulk actions}

In this appendix, we collect some basic background information and conventions
used throughout the paper. In computing moments, we encounter the
Green's function equation
\bb\label{greeneq}
\nabla_d^2 G=-(d-2) \Omega_{d-1} \delta^d(x)\Rightarrow G=\frac{1}{r^{d-2}}\ ,
\ee
where 
\bb
\Omega_{d-1}=\frac{2\pi^{d/2}}{\Gamma(d/2)}\ .
\ee
The linearized IIB supergravity action we use is written as 
\bb\label{sugraeq}
\LL=\frac{1}{2\kappa^2}\sqrt{|g|} \lk[
R+\frac{1}{2} g_s^2 (\del\chi)^2+\frac{1}{2} (\del\phi)^2+\frac{1}{3} \bar{F}_{abc} F^{abc}+\frac{1}{96} G_{abcde} G^{abcde}
\re]\ ,
\ee
where the gravitational coupling is
\bb
2\kappa^2=(2\pi)^7 {\alpha'}^{4}\ .
\ee
The 3-form flux is complex and has NSNS and RR pieces $H^{(1)}_{abc}$ and $H^{(2)}_{abc}$ respectively
\bb
F_{abc}=\frac{1}{2\sqrt{g_s}} H^{(1)}_{abc}+i\,\frac{\sqrt{g_s}}{2} H^{(2)}_{abc}\ ,
\ee
written to linear order in the fields.
We also write this in terms of the gauge fields
\bb
H^{(i)}=dA^{(i)}\ .
\ee
The self-dual 5-form field strength is written as
\bb
G=dA\ ,
\ee
with respect to a 4-form gauge field $A$ at infinity.
At the linearized order, no distinction is needed between tangent space and spacetime indices.
We use latin letters to label coordinates $a,b,c,\cdots$; these indices span over the eight
dimensional transverse subspace eluded to in the main text.

The Green-Schwarz worldsheet action in the light-cone gauge 
is~\cite{Sahakian:2004gy}
\bb\label{WSeq}
\II_{WS}=\int d^2\sigma\  \LL_s\ \ \ \mbox{with}\ \ \ \LL_s\equiv \LL^{(0)}+\LL^{(2)}+\LL^{(4)}
\ee
where the superscripts identify the number of spinors in each term. We have
\bb\label{I0}
\LL^{(0)}= 
-T \frac{\omega}{2}\sqrt{-h}\,h^{ij} V_i^a V_{a\, j}-2\, T\,\omega\,\sqrt{-h}\, h^{ij}\, V_i^+\, V_j^-
-\frac{T}{2} \varepsilon^{ij}\,V_i^a\,V_j^b\,A^{(1)}_{ab}
-2\,T\,\varepsilon^{ij} V^+_i\, V^-_j\, A^{(1)\,-+}\ ,
\ee
where the string tension is
\bb
T=\frac{1}{2\pi\alpha'}\ .
\ee
Note that the worldsheet is written in the Einstein frame, hence the additional $\omega\rightarrow\sqrt{g_s}$
dressing the tension. We also define
\bb
V_i^a\equiv \del_i x^m e_m^a\ \ \ ,\ \ \ V_i^\pm\equiv \del_i x^m e_m^\pm\ .
\ee
The term quadratic in the spinors is given by
\bb\label{I2}
\LL^{(2)}=i\, T\,\omega \sqrt{-h}\, h^{ij} V^+_i \overbar{\theta} \sigma^- D_j\theta
-i\,T\,\omega\, \varepsilon^{ij} V^+_i\,\theta\sigma^-D_j\theta-T\, V^{+\,i}\, V_{c}^j\,\II_{ij}^c+\mbox{c.c.}\ ,
\ee
where $\theta$ is a Weyl 16-component spinor and the $\sigma$'s are gamma matrices defined in
Appendix A.
We also have defined
\bbb\label{I2fields}
\II_{ij}^c&=&i\,\frac{\omega}{2}\,  \sqrt{-h}\, h_{ij}\, P^c\, \overbar{\theta}\sigma^- \theta
-i\,\frac{\omega}{2}\varepsilon_{ij}\, P_a\, \overbar{\theta}\sigma^{-c a}\overbar{\theta} \nonumber \\
&-&i \omega \sqrt{-h}\,h_{ij}\,\,F^{-+}_{\ \ a} \overbar{\theta}\sigma^{-ca}\overbar{\theta}
+i\frac{\omega}{4} \sqrt{-h}\,h_{ij}\,F_{ab}^{\ \ \ c} \overbar{\theta}\sigma^{-ab}\overbar{\theta} \nonumber \\
&+&i\omega \varepsilon_{ij} F^{-+c} \overbar{\theta}\sigma^-\theta
+i\omega \varepsilon_{ij} F^{-+}_{\ \ \ a} \overbar{\theta} \sigma^{-ca}\theta
+i \frac{\omega}{4}\,\varepsilon_{ij} F^c_{\ \ ab} \overbar{\theta}\sigma^{-ab} \theta\nonumber \\
&-& \frac{\omega}{4} \varepsilon_{ij} G^{-+c}_{\ \ \ \ ab} \theta \sigma^{-ab} \theta\ ,
\eee
where we see the couplings of the spinor bilinears on the worldsheet to the supergravity fields.
The covariant derivative appearing~\pref{I2} is written as
\bb\label{Dtheta}
D_j \theta\equiv \del_j \theta^\alpha 
-\frac{1}{4} \del_j x^m
\omega_{m,ab}\,\sigma^{ab}\theta
\ee
in terms of the spacetime connection $\omega_{m,ab}$.

We summarize the meaning of the various fields appearing in these expressions:
\begin{itemize}
\item The dilaton is written as
\bb
\omega\equiv e^{\phi/2}\ .
\ee

\item The field strength for the IIB scalars
\bb
P_a\equiv \frac{e^\phi}{2} \lk(i D_a \chi- e^{-\phi} D_a\phi\re)\ ;
\ee
with $\chi$ being the IIB axion.

\item The complex 3-form field strength 
\bb
F_{abc}\equiv
\frac{e^{\phi/2}}{2}
(1+e^{-\phi}+i\chi)\FF_{abc}
+\frac{e^{\phi/2}}{2}(-1+e^{-\phi}+i\chi)\bar{\FF}_{abc}
\ ;
\ee
with
\bb
\FF_{abc}\equiv \frac{H^{(1)}_{abc}}{2}+i\frac{H^{(2)}_{abc}}{2}\ ,
\ee
where $H^{(1)}$ and $H^{(2)}$ are, respectively, the field
strengths associated with the fundamental string and the D-string charges.

\item And the five-form self-dual field strength $G_{abcde}$.
\end{itemize}

The terms quartic in the spinors may also involve terms quadratic in the supergravity fields; we write
\bb\label{I4}
\LL^{(4)}=\sqrt{-h}\, h^{ij} V^+_i V^+_j \lk[\LL_{FF}+\LL_{FG}+\LL_{GG}+\LL_{DF}+\LL_{FP}+\LL_{DG}+\LL_R+\LL_{PP}\re]+\mbox{c.c.}
\ee
where we have defined the various pieces as:
\begin{eqnarray}
\LL_{GG}&=&-\frac{23\,\omega\, T}{4608}\,{\lk( \overbar{\theta}\,{\sigma }^{-}\,\theta \re) }^2\,\ 
{{G^{-\,+\,{a}\,{b}\,{c}}\ G^{-\,+}_{\ \ \ \,{a}\,{b}\,{c}}}\,
     }  \nonumber \\ &-& 
  \frac{\omega\, T}{9216}
  {\overbar{\theta}\,{\sigma }^{-\,{a}\,{b}}\,\theta }\ \,{\overbar{\theta}\,{\sigma }^{-}_{\ \ \,{a}\,{b}}\,\theta }\,\ 
  {{G^{-\,+\,{c}\,{d}\,{e}}\ G^{-\,+}_{\ \ \ \,{c}\,{d}\,{e}}}\,
     } 
     \nonumber \\ 
     &+& \frac{\omega\,T}{384} \overbar{\theta}\,{\sigma }^{-}\,\theta\  \,
      \overbar{\theta}\,{\sigma }^{-\,{a}\,{b}}\,\theta
      \left[
  \frac{1}{12}{{G}^{-\,+\,{c}\,{d}\,{e}}\,{G}_{{c}\,{d}\,{e}\,{a}\,{b}}\,
        } + 
 {{G}^{-\,+\,{c}}_{\ \ \ \ \ \ \,{a}\,{d}}\,
     {G}^{-\,+\,{d}}_{\ \ \ \ \ \ \,{\,{c}\,{b}}} \,
        } \right]\nonumber \\  
      &-& \frac{\omega\,T}{1536} \overbar{\theta}\,{\sigma }^{-\,{a}\,{b}}\,\theta\  \,
      \overbar{\theta}\,{\sigma }^{-\,{c}\,{d}}\,\theta 
  \left[{{G}^{-\,+\,{e}}_{\ \ \ \ \ \,{a}\,{b}}\,
     {G}^{-\,+}_{\ \ \ \,{e}\,{c}\,{d}} \,
        }  
     -  \frac{1}{24}{{G}_{{a}\,{b}\,{e}\,{f}\,{g}}\,{G}^{{e}\,{f}\,{g}}_{\ \ \ \ \,{c}\,{d}}} 
    \right] \nonumber \\ 
     &+& \frac{\omega\,T}{256} \overbar{\theta}\,{\sigma }^{-\,{a}\,{c}}\,\theta\  \,
      \overbar{\theta}\,{\sigma }^{-\,{b}}_{\ \ \ \ \,{c}}\,\theta 
      \left[
  {{G}^{-\,+}_{\ \ \ \ \,{a}\,{d}\,{e}}\,
     {G}^{-\,+\,{d}\,{e}}_{\ \ \ \ \ \ \ \,{b}}}-
\frac{1}{72}{{G}_{{a}\,{d}\,{e}\,{f}\,{g}}\,{G}^{{d}\,{e}\,{f}\,{g}}_{\ \ \ \ \ \ \,{b}} \,
       }\right]
\end{eqnarray}
\begin{eqnarray}
\LL_{FF}&=&
-\frac{13\,\omega\, T}{24}\,{\lk( \overbar{\theta}\,{\sigma }^{-}\,\theta \re) }^2\,\ 
{{F}^{-\,+\,{a}}\,{\overbar{F}}^{-\,+}_{\ \ \ \ \,{a}}} - 
  \frac{25\,\omega\, T}{768}\,{ \overbar{\theta}\,{\sigma }^{-\,{a}\,{b}}\,\theta }\ { \overbar{\theta}\,{\sigma }^{-}_{\ \,{a}\,{b}}\,\theta  }\,\ 
  {{F}^{-\,+\,{c}}\,{\overbar{F}}^{-\,+}_{\ \ \ \ \,{c}} \,
     } \nonumber \\ &-& \frac{\omega\, T}{256}\,
     \overbar{\theta}\,{\sigma }^{-}\,\theta\  \,
      \overbar{\theta}\,{\sigma }^{-\,{a}\,{b}}\,\theta  
      \left[
  93\,{{F}^{-\,+}_{\ \ \ \ \,{a}}\,{\overbar{F}}^{-\,+}_{\ \ \ \ \,{b}} \,
      } - 
  \frac{43}{2}\,{{F}^{-\,+\,{c}}\,{\overbar{F}}_{{c}\,{a}\,{b}} \,
      } - 
  \frac{17}{24}\,{{F}_{{a}\,{c}\,{d}}\,{\overbar{F}}^{{c}\,{d}}_{\ \ \ \,{b}} \,
      }\right] \nonumber \\ &+& \frac{\omega\, T}{96}\, 
  \overbar{\theta}\,{\sigma }^{-\,{a}\,{c}}\,\theta\  \,
      \overbar{\theta}\,{\sigma }^{-\,{b}}_{\ \ \ \ \,{c}}\,\theta  
      \left[17\,{{F}^{-\,+}_{\ \ \ \ \,{a}}\,{\overbar{F}}^{-\,+}_{\ \ \ \ \,{b}} \,
      }+ 
  \frac{5}{8}\,{{F}^{-\,+\,{d}}\,{\overbar{F}}_{{d}\,{a}\,{b}} \,
      } - 
  \frac{7}{16}\,{{F}_{{a}\,{d}\,{e}}\,{\overbar{F}}^{{d}\,{e}}_{\ \ \ \,{b}} \,
      }\right]
      \nonumber \\ &-& \frac{\omega\, T}{1536}\, 
  \overbar{\theta}\,{\sigma }^{-\,{a}\,{b}}\,\theta\  \,
      \overbar{\theta}\,{\sigma }^{-\,{c}\,{d}}\,\theta  
      \left[ 23\,{{F}^{-\,+}_{\ \ \ \ \,{d}}\,{\overbar{F}}_{{c}\,{a}\,{b}} \,
      }- 
  7\,{{F}^{-\,+}_{\ \ \ \ \,{a}}\,{\overbar{F}}_{{b}\,{c}\,{d}} \,
      } \right. \nonumber \\ &-& \left. 
  \frac{1}{2}{{F}_{{a}\,{c}\,{e}}\,{\overbar{F}}^{{e}}_{\ \,{b}\,{d}} \,
      }- 
\frac{13}{4}\,{{F}_{{a}\,{b}\,{e}}\,{\overbar{F}}^{{e}}_{\ \,{c}\,{d}} \,
      } \right]
  \end{eqnarray}
\begin{eqnarray}
\LL_{DG}&=&-\frac{\imag }{192}\,\omega\, T \,{D_{{c}}}\,G^{-\,+\,{c}}_{\ \ \ \ \ \,{a}\,{b}}  \,
    \overbar{\theta}\,{\sigma }^{-}\,\theta  \,
    \overbar{\theta}\,{\sigma }^{-\,{a}\,{b}}\,\theta 
\end{eqnarray}
\begin{eqnarray}
\LL_{PP}&=&-\frac{15\,\omega\, T}{256}\,\overbar{\theta}\,{\sigma }^{-}\,\theta  \,\ 
      \overbar{\theta}\,{\sigma }^{-\,{a}\,{b}}\,\theta\,\ 
      {{P}_{{a}}\,{\overbar{P}}_{{b}}} + 
  \frac{\omega\, T}{48}\,\overbar{\theta}\,{\sigma }^{-\,{a}\,{c}}\,\theta  \,\ 
      \overbar{\theta}\,{\sigma }^{-\,{b}}_{\ \ \ \ \,{c}}\,\theta\,\ 
      {{P}_{{a}}\,{\overbar{P}}_{{b}}}
\end{eqnarray}
\begin{eqnarray}
\LL_{R}&=&\frac{5\,\omega\, T}{32}\,\ 
{\lk( \overbar{\theta}\,{\sigma }^{-}\,\theta \re) }^2\, {R^{-\,+\,-\,+} \,
     }  -
  \frac{\omega\, T}{192}\,\ 
  {\overbar{\theta}\,{\sigma }^{-\,{a}\,{b}}\,\theta}\ {\overbar{\theta}\,{\sigma }^{-}_{\ \,{a}\,{b}}\,\theta}\,\ 
  {R^{-\,+\,-\,+}}  \nonumber \\ &+& \frac{\omega\, T}{96}\, 
     \overbar{\theta}\,{\sigma }^{-\,{a}\,{c}}\,\theta  \,\ 
      \overbar{\theta}\,{\sigma }^{-\,{b}}_{\ \ \ \,{c}}\,\theta  
      \left[
     {R^{-\,+}_{\ \ \ \ \,{a}\,{b}} \,
      } +
  \frac{1}{2}{R_{{a}\,{d}\,{b}}^{\ \ \ \ \,{d}}  
     }\right] \nonumber \\ &-& \frac{\omega\, T}{384}\, 
      \overbar{\theta}\,{\sigma }^{-\,{a}\,{b}}\,\theta  \,\ 
      \overbar{\theta}\,{\sigma }^{-\,{c}\,{d}}\,\theta  
       \left[
  {R_{{a}\,{c}\,{b}\,{d}} 
     }+
\frac{1}{2}{R_{{a}\,{b}\,{c}\,{d}}}\right] 
\end{eqnarray}
$R_{abcd}$ being the Riemann tensor in the Einstein frame.
And
\begin{eqnarray}
\LL_{FG}&=& -i \frac{\omega\, T }{48}\,\theta\,{\sigma }^{-\,{a}\,{b}}\,\theta\,\ \,\overbar{\theta}\,{\sigma }^{-}\,\theta 
\,{\overbar{F}}^{-\,+\,{e}}\,G^{-\,+}_{\ \ \ \,{a}\,{b}\,{e}}\,
     \,
       \nonumber \\
&-& i \frac{\omega\, T}{32}\,  \theta\,{\sigma }^{-\,{a}\,{b}}\,\theta\,\ \,
    \overbar{\theta}\,{\sigma }^{-\,{c}\,{d}}\,\theta
    \left[
    \frac{1 }{3}\,{\overbar{F}}_{{e}\,{b}\,{d}}\,G^{-\,+\,{e}}_{\ \ \ \ \ \,{a}\,{c}}\, \,
       -
  \frac{1 }{3}\,{\overbar{F}}^{-\,+}_{\ \ \ \ \,{d}}\,G^{-\,+}_{\ \ \ \,{a}\,{b}\,{c}}\,
       - 
  \frac{5}{12}\,{\overbar{F}}_{{e}\,{c}\,{d}}\,G^{-\,+\,{e}}_{\ \ \ \ \ \,{a}\,{b}}\, \right.   \nonumber \\ &-&\left.
  \frac{1}{12}\,{\overbar{F}}_{{e}\,{f}\,{d}}\,G^{{e}\,{f}}_{\ \ \,{a}\,{b}\,{c}}\,     + 
  \frac{1}{12}\,{\overbar{F}}_{{e}\,{a}\,{b}}\,G^{-\,+\,{e}}_{\ \ \ \ \ \,{c}\,{d}}\,    -
  \,{\overbar{F}}^{-\,+}_{\ \ \ \ \,{b}}\,G^{-\,+}_{\ \ \ \,{a}\,{c}\,{d}}\,
       + 
  \frac{1}{3}\,{\overbar{F}}^{-\,+}_{\ \ \ \ \,{e}}\,G^{{e}}_{\ \,{a}\,{b}\,{c}\,{d}}\,    \right] 
    \nonumber \\ 
    &-& i \frac{\omega\, T}{48}\, \theta\,{\sigma }^{-\,{a}\,{c}}\,\theta\,\ \,
    \overbar{\theta}\,{\sigma }^{-\,{b}}_{\ \ \ \ \,{c}}\,\theta
    \left[
  {\overbar{F}}^{-\,+}_{\ \ \ \ \,{d}}\,G^{-\,+\,{d}}_{\ \ \ \ \ \,{a}\,{b}}\,
       + \frac{1}{4}\,{\overbar{F}}_{{d}\,{e}\,{b}}\,G^{-\,+\,{d}\,{e}}_{\ \ \ \ \ \ \ \,{a}} \,
       -
  \frac{1}{4}\,{\overbar{F}}_{{d}\,{e}\,{a}}\,G^{-\,+\,{d}\,{e}}_{\ \ \ \ \ \ \ \,{b}} \,
      \right] \nonumber \\ &-& 
  i \frac{\omega\, T }{1152}\,
  \theta\,{\sigma }^{-\,{a}\,{b}}\,\theta\,\ \,
    \overbar{\theta}\,{\sigma }^{-}_{\ \ \,{a}\,{b}}\,\theta\,
    {\overbar{F}}_{{c}\,{d}\,{e}}\,G^{-\,+\,{c}\,{d}\,{e}} \, 
\end{eqnarray}
\begin{eqnarray}
\LL_{FP}&=&
-\frac{\omega\, T}{8}\,{{F}^{-\,+}_{\ \ \ \ \,{a}}\,{\overbar{P}}_{{b}}\ \,
      \overbar{\theta}\,{\sigma }^{-}\,\theta\  \,
      \theta\,{\sigma }^{-\,{a}\,{b}}\,\theta  } + 
  \frac{\omega\, T}{8}\,{{F}^{-\,+}_{\ \ \ \ \,{a}}\,{\overbar{P}}_{{b}}\ \,
      \overbar{\theta}\,{\sigma }^{-\,{a}\,{c}}\,\theta\  \,
      \theta\,{\sigma }^{-\,{b}}_{\ \ \ \,{c}}\,\theta  } \nonumber \\ &-& 
  \frac{\omega\, T}{96}\,{{F}_{{a}\,{c}\,{d}}\,{\overbar{P}}_{{b}}\ \,
      \overbar{\theta}\,{\sigma }^{-\,{c}\,{d}}\,\theta\  \,
      \theta\,{\sigma }^{-\,{a}\,{b}}\,\theta}
\end{eqnarray}
\begin{eqnarray}
\LL_{DF}&=&
-\frac{\omega\, T}{12} \, \theta\,{\sigma }^{-\,{a}\,{c}}\,\theta\,\ 
      \overbar{\theta}\,{\sigma }^{-\,{b}}_{\ \ \ \ \,{c}}\,\theta  \,\ 
      {{D_{{b}}}\,{\overbar{F}}^{-\,+}_{\ \ \ \ \,{a}}  \,
        } - 
  \frac{\omega\, T}{48} \, \theta\,{\sigma }^{-\,{a}\,{b}}\,\theta\,\ 
      \overbar{\theta}\,{\sigma }^{-\,{c}\,{d}}\,\theta  \, \ 
      {{D_{{c}}}\,{\overbar{F}}_{{a}\,{b}\,{d}}
      }\label{lasteq}
\end{eqnarray}

In computing moments, most of these quartic terms do not contribute.

\subsection{Appendix C: Spinor vevs}

In this appendix, we collect all vevs needed to compute the couplings on the worldsheet
for all possible 256 states of the state space of spinor zero modes. 

\vspace{0.2in}
{\bf Spinor bilears of the form $\bar{\theta}\theta$}

\bbb
\Bra{A,B} &\bar{\theta}\sigma^{-ab}\theta&\Ket{C,D}  \nonumber \\
&=& 2\,\Gamma \,\left( -{{\eta }^{A\,b}}\,{{\eta }^{B\,D}}\,{{\eta }^{a\,C }}  - 
    {{\eta }^{b\,B}}\,{{\eta }^{a\,D}}\,{{\eta }^{A\,C }} + {{\eta }^{a\,B}}\,{{\eta }^{b\,D}}\,{{\eta }^{A\,C }} + 
    {{\eta}^{a\,A}}\,{{\eta}^{B\,D}}\,{{\eta}^{b\,C }} \right)\ ; \nonumber \\
\Bra{A,B} \bar{\theta}\sigma^{-}\theta\Ket{C,D}  \nonumber
&=&4\,\Gamma \,{{\eta}^{B\,D}}\,{{\eta}^{A\,C }} \nonumber \\
\Bra{A,\alpha} \bar{\theta}\sigma^{-ab}\theta\Ket{C,\beta}  \nonumber 
&=&-\frac{\Gamma}{2}\left( 4\,{{\delta }_{\alpha \,\beta }}\,
         \left[ -{{\eta}^{A\,b}}\,{{\eta}^{a\,C }}   + {{\eta}^{a\,A}}\,{{\eta}^{b\,C }} \right]  + 
        { {\sigma }^{a\,b}_{\alpha \,\beta }}\,{{\eta}^{A\,C }} \right)\ ; \nonumber \\
\Bra{A,\alpha} \bar{\theta}\sigma^{-}\theta\Ket{C,\beta}  \nonumber
&=&-2\,\Gamma \,{{\delta }_{\alpha \,\beta }}\,{{\eta}^{A\,C }} \nonumber \\
\Bra{\alpha,\beta} \bar{\theta}\sigma^{-ab}\theta\Ket{\gamma,\omega}  \nonumber
&=&\frac{\Gamma}{2} \,\left( {{\delta }_{\beta \,\omega }}\,{{\sigma }^{a\,b}_{\alpha \,\gamma }} + 
      {{\delta }_{\alpha \,\gamma }}\,{{\sigma }^{a\,b}_{\beta \,\omega }} \right)\nonumber \\
\Bra{\alpha,\beta} \bar{\theta}\sigma^{-}\theta\Ket{\gamma,\omega}  \nonumber
&=&0\ .
\eee

\vspace{0.2in}
{\bf Spinor bilears of the form ${\theta}\theta$}

\bbb
\Bra{A,B} &{\theta}\sigma^{-ab}\theta&\Ket{C,D}  \nonumber \\
&=&2\,\Gamma \,\left( -{{\eta}^{A\,b}}\,{{\eta}^{B\,D}}\,{{\eta}^{a\,C }}  + 
    {{\eta}^{b\,B}}\,{{\eta}^{a\,D}}\,{{\eta}^{A\,C }} - {{\eta}^{a\,B}}\,{{\eta}^{b\,D}}\,{{\eta}^{A\,C }} + 
    {{\eta}^{a\,A}}\,{{\eta}^{B\,D}}\,{{\eta}^{b\,C }} \right)\nonumber \\
\Bra{A,\alpha} {\theta}\sigma^{-ab}\theta\Ket{C,\beta}  \nonumber 
&=&\frac{\Gamma}{2} \,\left( 4\,{{\delta }_{\alpha \,\beta }}\,\left[ {{\eta}^{A\,b}}\,{{\eta}^{a\,C }} - {{\eta}^{a\,A}}\,{{\eta}^{b\,C }} \right]
          + {{\eta}^{A\,C }}\,{{\sigma }^{a\,b}_{\alpha \,\beta }} \right) \nonumber \\
\Bra{\alpha,\beta} {\theta}\sigma^{-ab}\theta\Ket{\gamma,\omega}  \nonumber
&=&\frac{\Gamma}{2} \,\lk({{\delta }_{\beta \,\omega }}\,{{\sigma }^{a\,b}_{\alpha \,\gamma }} - 
    2 \,{{\delta }_{\alpha \,\gamma }}\,{{\sigma }^{a\,b}_{\beta \,\omega }}\re)\nonumber
\eee

\vspace{0.2in}
{\bf Terms quartic in the spinors of the form $\bar{\theta}\theta\,\bar{\theta}\theta$}

\vspace{0.2in}
Due to the ambiguity in ordering the bilinears in such quartic terms, we employ the
standard prescription of symmetrizing the bilinears before quantization; \ie\ 
we write for example
\bbb
:\bar{\theta}\sigma^{-ab}\theta\ \bar{\theta}\sigma^{-cd}\theta :=\frac{1}{2}\bar{\theta}\sigma^{-ab}\theta\ \bar{\theta}\sigma^{-cd}\theta+\frac{1}{2}\bar{\theta}\sigma^{-cd}\theta\ \bar{\theta}\sigma^{-ab}\theta\ .
\nonumber
\eee

\newpage
One then gets:
\bbb
\Bra{A,B} :{\overbar{\theta }}
    {\sigma }^{-ab}\theta \,
   {\overbar{\theta}}
    {\sigma }^{-cd}\theta : \Ket{C,D}&=& 
  8{\Gamma }^2\left( {{\eta}^{Bc}}{{\eta}^{bC}}
      {{\eta}^{Ad}}{{\eta}^{aD}} + 
     {{\eta}^{Bc}}{{\eta}^{AC}}{{\eta}^{bd}}
      {{\eta}^{aD}} + {{\eta}^{Ab}}{{\eta}^{cC}}
      {{\eta}^{Bd}}{{\eta}^{aD}} \right. \nonumber \\
    &+&  \left. {{\eta}^{bB}}{{\eta}^{Ac}}{{\eta}^{Cd}}
      {{\eta}^{aD}} + {{\eta}^{AB}}{{\eta}^{bc}}
      {{\eta}^{Cd}}{{\eta}^{aD}} 
      +  {{\eta}^{Bc}}{{\eta}^{aC}}
      {{\eta}^{bd}}{{\eta}^{AD}}   \right. \nonumber \\
      &+&  \left.{{\eta}^{aB}}{{\eta}^{cC}}{{\eta}^{bd}}
      {{\eta}^{AD}} + {{\eta}^{ac}}{{\eta}^{BC}}
      {{\eta}^{Ad}}{{\eta}^{bD}} 
      + {{\eta}^{ac}}{{\eta}^{bC}}{{\eta}^{Ad}}
      {{\eta}^{BD}}\right. \nonumber \\
      &+& \left. 
     {{\eta}^{aA}}{{\eta}^{cC}}{{\eta}^{bd}}
      {{\eta}^{BD}} + {{\eta}^{aA}}{{\eta}^{BC}}
      {{\eta}^{bd}}{{\eta}^{cD}} + 
     {{\eta}^{bB}}{{\eta}^{Ac}}{{\eta}^{ad}}
      {{\eta}^{CD}}\right. \nonumber \\
      &+& \left.
     {{\eta}^{aA}}{{\eta}^{bc}}{{\eta}^{Bd}}
      {{\eta}^{CD}} + {{\eta}^{AB}}{{\eta}^{bc}}
      {{\eta}^{aC}}{{\eta}^{dD}} + 
     {{\eta}^{Ab}}{{\eta}^{Bc}}{{\eta}^{aC}}
      {{\eta}^{dD}} \right. \nonumber \\ 
      &+& \left.  {{\eta}^{bB}}{{\eta}^{ac}}
      {{\eta}^{AC}}{{\eta}^{dD}} 
      +{{\eta}^{aB}}{{\eta}^{Ac}}{{\eta}^{bC}}
      {{\eta}^{dD}} + {{\eta}^{aA}}{{\eta}^{bB}}
      {{\eta}^{cC}}{{\eta}^{dD}}\right. \nonumber \\
            &+& \left. 
     2{{\eta}^{bc}}{{\eta}^{BC}}{{\eta}^{ad}}
      {{\eta}^{AD}} + 2{{\eta}^{AB}}
      {{\eta}^{ac}}{{\eta}^{bd}}{{\eta}^{CD}}  + 2{{\eta}^{ac}}
      {{\eta}^{AC}}{{\eta}^{bd}}{{\eta}^{BD}} 
 	\right)\left.\right|_{[a,b],[c,d]} \nonumber
\eee
The notation $|_{[a,b],[c,d]}$ signifies antisymmetrization of the enclosed expression for $[a,b]$
and $[c,d]$ separately.

\bbb
\Bra{A,\alpha}:{\overbar{\theta}}
    {\sigma }^{-ab}\theta \,
   {\overbar{\theta}}
    {\sigma }^{-cd}\theta : \Ket{C,\beta}&=&
  {\Gamma }^2\left( -8{{\delta }_{\alpha \beta }}
      {{\eta}^{Ab}}{{\eta}^{cC}}{{\eta}^{ad}} + 
     8{{\delta }_{\alpha \beta }}{{\eta}^{ac}}
      {{\eta}^{bC}}{{\eta}^{Ad}} \right. \nonumber \\ 
        &+& \left. 
     \frac{3}{2}{{\eta}^{cC}}{{\eta}^{Ad}}
        {{\sigma }^
             {-ab}_
          {\alpha \beta }}  + 
     \frac{3}{2}{{\eta}^{Ab}}{{\eta}^{aC}}
        {{\sigma }^
             {-cd}_{\alpha \beta }} 
        +
     \frac{1}{2}{{\eta}^{ac}}{{\eta}^{bd}}
        {{\sigma }^
             {-AC}_{\alpha \beta }} - 
     \frac{1}{2}{{\eta}^{AC}}
        {{\sigma }^
             {-abcd}_{\alpha \beta }}  \right. \nonumber \\ 
                   &+& \left. 
     {{\eta}^{cC}}{{\eta}^{bd}}
      { {\sigma }^
           {-aA}_
        {\alpha \beta}}+ 
     {{\eta}^{bC}}{{\eta}^{ad}}
      {{\sigma }^
           {-Ac}_
        {\alpha \beta}} \right. \nonumber \\
        &+& \left.
     {{\eta}^{Ab}}{{\eta}^{ad}}
      {{\sigma }^
           {-cC}_{\alpha \beta }} + 
     {{\eta}^{bc}}{{\eta}^{Ad}}
      {{\sigma }^
           {-aC}_
        {\alpha \beta }}  + 
     {{\eta}^{aC}}{{\eta}^{Ad}}
      {{\sigma }^
           {-bc}_
        {\alpha \beta }} - 
     {{\eta}^{aA}}{{\eta}^{Cd}}
      {{\sigma }^
           {-bc}_
        {\alpha \beta }}
     \right)\left.\right|_{[a,b],[c,d]} \nonumber
\eee

\bbb
\Bra{\alpha,\beta}:{\overbar{\theta}}
    {\sigma }^{-ab}\theta \,
   {\overbar{\theta}}
    {\sigma }^{-cd}\theta : \Ket{\gamma,\omega}&=&
  \frac{{\Gamma }^2}{2}\left( -\frac{1}{2} {{\sigma }^
                 {-ab}_
              {\beta \omega }}                     
            {{\sigma }^
                 {-cd}_
              {\alpha \gamma }}
           -
                   \frac{1}{2}{{\sigma }^
               {-ab}_
            {\alpha \gamma }}
          {{\sigma }^
               {-cd}_
            {\beta \omega }}\right. \nonumber \\
             &-&  \left.
       {{\delta }_{\beta \omega }}
        {{\sigma }^
             {-abcd}_
          {\alpha \gamma }} - 
       {{\delta }_{\alpha \gamma }}
        {{\sigma }^
             {-abcd}_
          {\beta \omega }} \right)\nonumber
\eee

\vspace{0.2in}
{\bf Terms quartic in the spinors of the form ${\theta}\theta\,\bar{\theta}\theta$ or c.c.}

\bbb
\Bra{A,B}:\theta 
    {\sigma }^{-ab}\theta \,
   {\overbar{\theta}}
    {\sigma }^{-cd}\theta : \Ket{C,D} &=& 
  8{\Gamma }^2\left( {{\eta}^{bc}}{{\eta}^{AC}}
      {{\eta}^{Bd}}{{\eta}^{aD}} + 
     2{{\eta}^{aB}}{{\eta}^{Ac}}{{\eta}^{Cd}}
      {{\eta}^{bD}} \right. \nonumber \\ 
      &+& \left.{{\eta}^{ac}}{{\eta}^{bC}}
      {{\eta}^{Ad}}{{\eta}^{BD}} + 
     {{\eta}^{Ab}}{{\eta}^{ac}}{{\eta}^{Cd}}
      {{\eta}^{BD}} + 2{{\eta}^{Ab}}
      {{\eta}^{Bc}}{{\eta}^{aC}}{{\eta}^{dD}} \right. \nonumber \\
      &+& \left.  
     {{\eta}^{aB}}{{\eta}^{bc}}{{\eta}^{AC}}
      {{\eta}^{dD}} \right)\nonumber
\eee

\bbb
\Bra{A,\alpha}:\theta
   {\sigma }^{-ab}\theta \,
  {\overbar{\theta}}
   {\sigma }^{-cd}\theta : \Ket{C,\beta}=0\nonumber
\eee

\bbb
 \Bra{\gamma,\omega}:\theta
   {\sigma }^{-ab}\theta \,
  {{\theta }^{\_}}
   {\sigma }^{-cd}\theta :  \Ket{\gamma,\omega}=0\nonumber
\eee
   

\begin{thebibliography}{10}

\bibitem{Emparan:2001wn}
R.~Emparan and H.~S. Reall,
\newblock Phys. Rev. Lett. {\bf 88}, 101101 (2002), hep-th/0110260.

\bibitem{Elvang:2004rt}
H.~Elvang, R.~Emparan, D.~Mateos, and H.~S. Reall,
\newblock Phys. Rev. Lett. {\bf 93}, 211302 (2004), hep-th/0407065.

\bibitem{Bena:2004de}
I.~Bena and N.~P. Warner,
\newblock (2004), hep-th/0408106.

\bibitem{Gauntlett:2004qy}
J.~P. Gauntlett and J.~B. Gutowski,
\newblock Phys. Rev. {\bf D71}, 045002 (2005), hep-th/0408122.

\bibitem{Bena:2005ni}
I.~Bena, P.~Kraus, and N.~P. Warner,
\newblock Phys. Rev. {\bf D72}, 084019 (2005), hep-th/0504142.

\bibitem{Bena:2005ay}
I.~Bena and P.~Kraus,
\newblock Phys. Rev. {\bf D72}, 025007 (2005), hep-th/0503053.

\bibitem{Bena:2005va}
I.~Bena and N.~P. Warner,
\newblock (2005), hep-th/0505166.

\bibitem{Jejjala:2005yu}
V.~Jejjala, O.~Madden, S.~F. Ross, and G.~Titchener,
\newblock Phys. Rev. {\bf D71}, 124030 (2005), hep-th/0504181.

\bibitem{Saxena:2005uk}
A.~Saxena, G.~Potvin, S.~Giusto, and A.~W. Peet,
\newblock (2005), hep-th/0509214.

\bibitem{Mathur:2005zp}
S.~D. Mathur,
\newblock Fortsch. Phys. {\bf 53}, 793 (2005), hep-th/0502050.

\bibitem{Mathur:2005ai}
S.~D. Mathur,
\newblock (2005), hep-th/0510180.

\bibitem{Lunin:2001dt}
O.~Lunin and S.~D. Mathur,
\newblock Nucl. Phys. {\bf B615}, 285 (2001), hep-th/0107113.

\bibitem{Lunin:2001jy}
O.~Lunin and S.~D. Mathur,
\newblock Nucl. Phys. {\bf B623}, 342 (2002), hep-th/0109154.

\bibitem{Bena:2004wv}
I.~Bena,
\newblock Phys. Rev. {\bf D70}, 105018 (2004), hep-th/0404073.

\bibitem{Bena:2004wt}
I.~Bena and P.~Kraus,
\newblock Phys. Rev. {\bf D70}, 046003 (2004), hep-th/0402144.

\bibitem{Plefka:1997xq}
J.~Plefka and A.~Waldron,
\newblock Nucl. Phys. {\bf B512}, 460 (1998), hep-th/9710104.

\bibitem{Millar:2000ib}
K.~Millar, W.~Taylor, and M.~Van~Raamsdonk,
\newblock (2000), hep-th/0007157.

\bibitem{Sahakian:2004gy}
V.~Sahakian,
\newblock JHEP {\bf 04}, 026 (2004), hep-th/0402037.

\bibitem{Taylor:2005db}
M.~Taylor,
\newblock (2005), hep-th/0507223.

\bibitem{Banks:1996vh}
T.~Banks, W.~Fischler, S.~H. Shenker, and L.~Susskind,
\newblock Phys. Rev. {\bf D55}, 5112 (1997), hep-th/9610043.

\bibitem{Dijkgraaf:1997vv}
R.~Dijkgraaf, E.~Verlinde, and H.~Verlinde,
\newblock Nucl. Phys. {\bf B500}, 43 (1997), hep-th/9703030.

\bibitem{Myers:1999ps}
R.~C. Myers,
\newblock JHEP {\bf 12}, 022 (1999), hep-th/9910053.

\bibitem{Howe:1983sr}
P.~S. Howe and P.~C. West,
\newblock Nucl. Phys. {\bf B238}, 181 (1984).

\end{thebibliography}

\end{document}